\documentclass[usenatbib]{mn2e}
\usepackage{natbib}
\usepackage{epsf}
\usepackage{graphicx}
\usepackage{hyperref}

\title[Multi-wavelength polarization of Cyg X--1]{The multi-wavelength polarization of Cygnus X--1}
\author[D. M. Russell \& T. Shahbaz]
{David M. Russell$^{1,2,3}$\thanks{E-mail: dave.russell@nyu.edu} and Tariq Shahbaz$^{1,2}$
\\
$^1$Instituto de Astrof\'isica de Canarias (IAC), E-38200 La Laguna, Tenerife, Spain \\ 
$^2$Departamento de Astrof\'isica, Universidad de La Laguna (ULL), E-38206 La Laguna, Tenerife, Spain \\
$^3$New York University Abu Dhabi, P.O. Box 129188, Abu Dhabi, United Arab Emirates \\
}

\def\simlt{\mathrel{\rlap{\lower 3pt\hbox{$\sim$}}
        \raise 2.0pt\hbox{$<$}}}
\def\simgt{\mathrel{\rlap{\lower 3pt\hbox{$\sim$}}
        \raise 2.0pt\hbox{$>$}}}

\begin{document}
\maketitle

\begin{abstract}
Polarization measurements of the microquasar Cygnus X--1 exist at $\gamma$-ray, X-ray, UV, optical and radio frequencies. The $\gamma$-ray emission has been shown to be highly linearly polarized. Here, we present new infrared polarimetric data of Cygnus X--1 taken with the 10.4-m Gran Telescopio Canarias and the 4.2-m William Herschel Telescope. We show that the broadband, radio to $\gamma$-ray flux spectrum and polarization spectrum in the hard state are largely consistent with a simple phenomenological model of a strongly polarized synchrotron jet, an unpolarized Comptonized corona and a moderately polarized interstellar dust component. In this model, the origin of the $\gamma$-ray, X-ray and some of the infrared polarization is the optically thin synchrotron power law from the inner regions of the jet. The model requires the magnetic field in this region to be highly ordered and perpendicular to the axis of the resolved radio jet. This differs to studies of some other X-ray binaries, in which the magnetic field is turbulent, variable and aligned with the jet axis. The model is able to explain the approximate polarization strength and position angle at all wavelengths including the detected X-ray (3--5 keV) polarization, except the observed position angle of the $\gamma$-ray polarization, which differs to the model by $\sim 60^{\circ}$. Past numerical modelling has shown that a curved synchrotron spectrum can produce a shift in position angle by $\sim 60^{\circ}$, which may account for this.
\end{abstract}


\section{Introduction}

X-ray binaries are binary systems in which a compact object, a black hole or a neutron star, accretes matter from a companion star. The polarization properties of emission from black hole X-ray binaries (BHXBs) have been well studied at radio frequencies \citep[see][ for a review]{fend06}, and have recently been given some attention at optical and infrared (IR) frequencies. In the optical regime, polarization due to the scattering of intrinsically unpolarized thermal emission can be modulated on the orbital period, which places constraints on the physical and geometrical properties of the system \citep{dolata89,glioet98}. At radio, and in some cases at optical/IR frequencies, variable polarization can be due to synchrotron emission from the jets launched via the process of accretion onto the black hole or neutron star in X-ray binaries \citep{hannet00,brocet07,shahet08,russfe08,russet11b}.

More than thirty years ago, a few measurements of polarization from X-ray binaries were made at X-ray energies using the Bragg crystal polarimeters aboard the OSO 8 satellite. The linear polarization of the high-mass X-ray binary (HMXB) and microquasar, Cygnus X--1 was measured to be 2--5 per cent at 2.5--5.2 keV \citep*[the confidence level was 3.9 $\sigma$ at 2.6 keV;][]{longet80}. Since then, no X-ray detector has had the capabilities to measure the polarization properties of X-ray binaries more accurately. More recently, the INTEGRAL satellite has been used, using novel techniques, to estimate the polarization of the Crab \citep{deanet08}, Cygnus X--1 \citep{lauret11,jouret12} and some $\gamma$-ray bursts \citep[GRBs;][]{gotzet09,gotzet13} at hard X-ray--$\gamma$-ray energies. The $\gamma$-ray polarization of Cyg X--1 was found to be very high, $67 \pm 30$ per cent at 0.4--2 MeV using the IBIS instrument on board INTEGRAL \citep{lauret11}, later confirmed using the SPI instrument, $76 \pm 15$ per cent at 0.23--0.85 MeV \citep{jouret12}. The only viable mechanism for producing such high polarization at these energies is optically thin synchrotron emission, and it was claimed that the high energy electrons in the jet are the origin of the polarization \citep{lauret11,jouret12}. Detailed spectral modelling of the keV--MeV \citep*[e.g.][]{mccoet02,zdziet12,delset13} and keV--GeV \citep*{malyet13,zdziet13} emission of Cyg X--1 has confirmed the presence of an MeV tail in the hard state, which could either be due to hybrid Comptonization or a synchrotron component. Here we adopt the state classifications of \cite{bell10}. If the high levels of polarization at 0.2--2 MeV are robust, synchrotron is the favoured mechanism, and jet models are consistent with the MeV tail being the high energy extension of the optically thin synchrotron power law extending from infrared wavelengths \citep{rahoet11,zdziet12,zdziet13,malyet13}. Alternatively, it has been suggested that a hot accretion flow could also produce synchrotron emission that is highly polarized at MeV energies; this requires one-dimensional motion of electrons along highly ordered magnetic field lines in the inner regions of the hot flow \citep*{veleet13}. In this case, a small region of the hot flow accreting from a preferential direction would presumably have to dominate the MeV emission, because the field lines threading the flow from all parts of the inner accretion disc would have different orientations.

Optically thin synchrotron emission is intrinsically polarized. If the local magnetic field in the emitting region is uniform (ordered), a net linear polarization is observed. If the field is tangled, the differing angles of the polarized light suppress the observed, average polarization. The maximum polarization strength is 
$\sim 70$--80 per cent, in the case of a perfectly ordered field \citep[e.g.][]{rybili79,bjorbl82} and is dependent only on the degree of ordering of the field and the energy distribution of the electron population (see also Section 3). When the radio emission of BHXBs is consistent with optically thin synchrotron (this typically occurs during X-ray state transitions), this polarization signature is commonly detected at relatively high levels; $\sim 10-30$ per cent \citep[e.g.][]{hannet00,fendet02,brocet07,robeet08,millet08,curret13}, and the emission here is from discrete jet ejections or interactions with the interstellar medium (ISM) and are often resolved in radio images. The position angle (PA) of polarization in these ejections is often (but not always) approximately parallel to the axis of the resolved radio jet, implying that the electric field is parallel to the jet axis and the magnetic field is orthogonal to the jet axis. This may be due to the compression of tangled magnetic field lines in shocks downstream in the flow or collisions with dense regions of the ISM, resulting in a partially ordered transverse field.

Compact, conical jets are known to produce a flat or slightly inverted spectrum from radio to infrared frequencies in BHXBs \citep[$0 \leq \alpha_{\rm thick} \leq +0.5$, where $F_{\nu}\propto \nu^{\alpha}$;][]{fendet00,fendet01,corbfe02,miglet07,gallet07,rahoet11,gandet11,russet13b}. These continuously-launched jets are different from the discrete ejections; their spectra are composed of overlapping, self-absorbed (optically thick) synchrotron components originating from distributions of electrons with various energies propagating down the jet, much like those of active galactic nuclei \citep[AGN;][]{blanko79,kais06}. In BHXBs these compact jets are produced when the source is in the hard X-ray state \citep*[e.g.][]{gallet03}. Linear polarization has been detected from this optically thick flat spectrum at a level of a few per cent at radio frequencies in a few BHXBs \citep[e.g.][]{gallet04,brocet13}. At some frequency, generally considered to lie in the infrared regime, this synchrotron spectrum breaks to one which is optically thin, with $-1 \leq \alpha_{\rm thin} \leq -0.4$. The power law spectrum of the optically thin emission has been identified and isolated in several BHXBs in the infrared/optical \citep*[e.g.][]{hyneet03,hyneet06,homaet05,kaleet05,chatet11} and the break itself has been detected in a few BHXBs in the mid-infrared \citep{corbfe02,rahoet11,gandet11,russet13a}.

To date, few studies have attempted to uncover the polarimetric signature of the optically thin synchrotron emission from compact jets that exist in the hard state in BHXBs. This emission originates close to the base of the jet, in a region likely associated with the start of the particle acceleration in the jet \citep*[e.g.][]{polket10}. The polarization seen from this region could have a higher level of ordering compared to further out in the jet, since the field may maintain a high level of ordering over the smaller emission region \citep{blanko79}. Polarimetric measurements of the optically thin power law therefore provide a powerful tool to uncover the nature of the magnetic field structure in this region, which is important for models and simulations of jet production. In the optical/IR regime of X-ray binaries, other components such as the accretion disc and companion star often dominate, suppressing any synchrotron contribution to the polarization, but when the synchrotron makes a strong contribution, intrinsic polarization has been detected \citep[e.g.][]{dubuch06,shahet08,russfe08,russet11b,chatet11}. The fractional linear polarization (FLP) is on the order of $\sim 1$--10 per cent, with evidence for rapid variations in some sources on timescales of seconds--minutes. The PA is usually approximately orthogonal to the axis of the resolved radio jet when this angle is known, which implies the magnetic field is parallel to the jet axis. The observations to date are consistent with a variable, predominantly tangled magnetic field geometry, with field lines preferentially orientated along the jet axis.

Here, we present new, high-precision NIR polarization measurements of Cyg X--1, a persistently active X-ray binary that is known to launch a powerful jet \citep{gallet05,russet07}. We gather archival flux spectral energy distributions (SEDs) and all polarization measurements of the source published to date, and attempt to model the multi-wavelength flux spectrum, FLP spectrum and PA spectrum self-consistently. Section 2 describes the data collection and treatment, and the model and results are presented in Section 3. A discussion is provided in Section 4, including predictions for future X-ray polarization detections of X-ray binaries. The conclusions are summarised in Section 5.

\begin{table}
\caption{Log of GTC observations. All observations are at 10.3\,$\mu$m (Si-4 filter).}
\vspace{-8mm}
\begin{center}
\begin{tabular}{lcccl}\hline
Target        & UT Date start  & exp. time &  PWV$^a$ & Nature \\
\hline
HD\,184827  & 2013-08-06  02:18 & 73\,s           & 12.8 & zero pol.   \\
Cyg\,X--1   & 2013-08-06  02:54 & 3$\times$3$\times$291\,s & 13.1 & OB \# 1    \\
MWC\,349    & 2013-08-06  04:06 & 73\,s           & 12.0 & polarized  \\ \\
HD\,184827  & 2013-10-04  23:55 & 73\,s           & 4.2  & zero pol.   \\
Cyg\,X--1   & 2013-10-05  00:19 & 3$\times$3$\times$291\,s & 5.2  & OB \# 2    \\
MWC\,349    & 2013-10-05  01:32 & 73\,s           & 6.8  & polarized  \\ \\
Cyg\,X--1   & 2013-10-05  22:53 & 3$\times$3$\times$291\,s & 6.1  & OB \# 3    \\
HD\,184827  & 2013-10-06  00:11 & 73\,s           & 4.6  & zero pol.   \\
MWC\,349        & 2013-10-06  00:37 & 73\,s & 6.4  & polarized  \\ \hline
\end{tabular}
\end{center}
$^a$PWV is the Precipitable Water Vapour (mm) \\
\label{table:gtc_log}
\end{table}

\begin{table}
\caption{Cyg X--1 GTC polarization measurements.}
\vspace{-2mm}
\begin{center}
\begin{tabular}{lccc}\hline
OB      &      Stokes $q$          &       Stokes $u$ & Flux (Jy) \\ \hline
\# 1    &  0.0101  $\pm$ 0.0258  & -0.0126  $\pm$ 0.0276 & 1.073$^a$\\
\# 2    &  0.0203  $\pm$ 0.0723  & -0.0093  $\pm$ 0.0601 & 0.34  \\
\# 3    & -0.0186  $\pm$ 0.0367  &  0.0002  $\pm$ 0.0350 & 0.33 \\
Average &  0.0039  $\pm$ 0.0284  & -0.0072  $\pm$ 0.0249 & \\
\hline
\end{tabular}
\end{center}
$^a$This flux is likely to be inaccurate due to the worse PWV value on this date.
\label{table:gtc_pol}
\end{table}

\begin{table}
\caption{Log of WHT observations. All observations are of Cyg X--1.}
\vspace{-8mm}
\begin{center}
\begin{tabular}{lcccl}\hline
Date & UT start--end  & Airmass & Filters & Mode$^a$ \\
\hline
2010-06-18 & 03:51--04:24 & $\leq 1.03$ & $J$, $H$, $K_{\rm S}$ & ROT \\
2013-09-13 & 02:35--04:10 & 2.1--5.0 & $J$, $H$, $K_{\rm S}$ & HWP \\
2013-09-15 & 02:28--03:05 & 2.1--2.7 & $Z$ & HWP \\
\end{tabular}
\end{center}
$^a$ROT is the method of rotating the camera by $90^{\circ}$, HWP is the method of using the LIRIS half-wave plate. \\
\label{table:wht_log}
\end{table}

\section{Data collection}

\subsection{Mid-IR polarization observations with the Gran Telescopio Canarias}

CanariCam polarimetric observations of Cyg X--1 were taken during 2013 August and October with the Gran Telescopio Canarias (GTC) on La Palma. We used the  Wollaston prism, half wave retarder (half-wave plate, HWP) and the Silicate filter Si-4 centred at 10.3\,$\mu$m (bandwidth 0.9 $\mu$m) to obtain  dual beam linear polarimetry of our targets. We observed Cyg X--1 for a total of 131\,min  with a chop angle of 90 degrees, chop throw of 8 arcsecs, nod angle $= -90$ degrees and a nod throw of 8 arcsecs. To measure the instrumental polarization  an  unpolarized standard  star HD\,184827 was observed. We also observed the polarized star MWC\,349 in order to determine the position angle offset. These calibration observations were taken every time Cyg X--1 was observed.
The amount of Precipitable Water Vapour (PWV) in the atmosphere in terms of millimeters was measured by the IAC real-time PWV monitor during each observation. The October observations were taken under much better PWV conditions compared to the  August observations (see Table\,\ref{table:gtc_log} for a log of observations).

In polarimetry mode the HWP rotates automatically between the four position angles, 0$^{\circ}$, 45$^{\circ}$, 22.5$^{\circ}$ and 67.5$^{\circ}$. The rotation of the HWP is synchronised with the chopping and nodding so that the final raw image cube contains several extensions, each corresponding to a wave plate angle. The Stokes parameters for a source are determined from a combination of the ordinary and extraordinary images for each value of the HWP angle. The data were reduced using an automated set of \textsc{pyraf} scripts specifically created to measure the polarization of point sources, provided by the GTC science operations team.  \textsc{pyraf} is a language for running \textsc{iraf}\footnote{\textsc{iraf} is distributed by the National Optical Astronomy Observatory, which is operated by the Association of Universities for Research in Astronomy, Inc., under cooperative agreement with the National Science Foundation. \url{http://iraf.noao.edu/}} tasks that is based on the \textsc{python} scripting language \citep{greewh00}. Aperture photometry was performed on the target and calibration stars using a fixed aperture of radius 5 pixels. The Stokes parameters were determined using the formulae described in  \citet{tin05}. The results of these scripts were cross-checked with \textsc{Starlink}\footnote{\url{http://starlink.jach.hawaii.edu/}} package \textsc{POLPACK}, which was producing the same results within the errors.

The instrumental Stokes $q$ and $u$ values determined using the non-polarized star were subtracted from the individual Cyg\,X--1 $q$ and $u$ values on each date. These corrected values from the three dates were then combined to give  $q = 0.0039 \pm 0.0284$ and $u = -0.0072 \pm 0.0249$ which equates to a polarization of $FLP = 0.82 \pm 2.57$ per cent and a position angle of $149 \pm 96^{\circ}$ (see Table\,\ref{table:gtc_pol} for polarization results). The $3 \sigma$ upper limit is $FLP < 8.53$ per cent.

For the three observations of Cyg X--1 we measured the total intensity ratio with respect to the mid-IR standard star HD184827 \citep{coheet92}. The flux density in the Si-4 filter (10.3 $\mu$m) of the standard is 10.925 Jy, and the flux densities of Cyg X--1 were 1.073 Jy, 0.34 Jy and 0.33 Jy for the
three observing blocks (OBs) \#1, \#2, \#3,  respectively. The flux density is very similar for observing blocks \#2 and \#3, but differs by a factor of 3 for observing block \#1, which is very likely due to the much worse PWV on this date (see Table\,\ref{table:gtc_log}). We therefore discard this flux density value from the following analysis.

\begin{table}
\begin{center}
\caption{Summary of multi-wavelength fluxes collected of Cyg X--1 in the hard state (flux upper limits are not included).}
\vspace{-5mm}
\begin{tabular}{lll}
\hline
Waveband&$log$($\nu$; Hz)&Reference\\
\hline
235--610 MHz&8.37--8.79&\cite{pandet07}\\
2.3--221 GHz&9.36--11.34&\cite{fendet00}\\
5.5--27 $\mu$m$^{a}$&13.05--13.73&\cite{rahoet11}$^{b}$\\
5--18 $\mu$m&13.22--13.78&\cite{miraet96}\\
2.3--10 $\mu$m&13.48--14.12&\cite{perset80}\\
1.2--2.2 $\mu$m&14.13--14.39&\cite{skruet06}\\
1.2--2.2 $\mu$m&14.13--14.39&This paper\\
0.44--0.55 $\mu$m&14.74--14.83&\cite{brocet99}\\
0.37 $\mu$m&14.91&\cite{breget73}\\
0.122--0.56 $\mu$m$^{a,c}$&14.73--15.39&\cite{cabaet09}\\
3.5--160 keV$^{a}$&17.93--19.59&\cite{rahoet11}$^{b}$\\
260--5400 keV$^{a}$&19.80--21.12&\cite{zdziet12}\\
0.1--10 GeV$^{a}$&22.38--24.37&\cite{malyet13}\\
\hline
\end{tabular}
\end{center}
$^a$Includes spectroscopic data. $^b$Data from Spitzer observation 1 as defined in \cite{rahoet11} were used, as the source was in the hard state and the jet was present. $^c$The UV data were taken when the source was in a soft state.
\end{table}

\subsection{NIR polarization observations with the William Herschel Telescope}

We observed Cyg X--1 with the Long-slit Intermediate Resolution Infrared Spectrograph (LIRIS) in imaging polarimetry mode on the 4.2-m William Herschel Telescope (WHT) at the Observatorio del Roque de los Muchachos, La Palma, Spain. The data were taken on 2010 June 18, 2013 September 13 and 15 (see Table \ref{table:wht_log} for the log of observations). Conditions were good on all dates, with some thin cirrus only on 2013 September 13. In 2010 the airmass was excellent, $\leq 1.03$, and in 2013 the airmass varied between 2.1 and 5.0. Exposures were made in a five-point dither pattern, separately in $Z$, $J$, $H$ and $K_{\rm S}$ filters, and a neutral density filter was used due to the high brightness of the source. The Wollaston prism splits the incoming light into four simultaneous images, one at each of the four polarization angles; 0$^{\circ}$, 45$^{\circ}$, 90$^{\circ}$ and 135$^{\circ}$. For the 2010 observations, half of the observations were made with the telescope rotator at $0^{\circ}$ and half at $90^{\circ}$, in order to correct for the relative transmission factors of the ordinary and extraordinary rays for each Wollaston \citep[see][]{alveet11,zapaet11}. In 2013, we made use of the new, achromatic half-wave plate recently available on LIRIS. The use of the half-wave plate ensures camera rotation is no longer necessary, and saves observing time since camera rotation significantly increases overheads.

The data reduction was performed using the \textsc{lirisdr} package developed by the LIRIS team in the \textsc{iraf} environment \citep[for details, see][]{alveet11}. Aperture photometry was then performed on the resulting combined images, and the normalized Stokes parameters $q$ and $u$, and FLP and PA were measured using equations (11--13) in \cite{alveet11} for the 2010 data. Errors on FLP and PA were computed using a Monte Carlo routine that propagates the errors associated with the raw counts at each polarization angle. For the 2013 data, equations presented in Pereyra \& Acosta-Pulido (in preparation) that apply to half-wave plate data were adopted (almost identical equations, except for sign changes in $q$ and $u$). The instrumental polarization is known to be very small for LIRIS; $< 0.1$ per cent \citep{alveet11}. However, in $J$ and $H$-bands, in which the FLP agrees very well with that expected from interstellar dust in the 2010 data (see Section 3.3), we found that the measured PA was offset from the known optical PA of polarization due to interstellar dust by $(3.19 \pm 0.73)^{\circ}$ for the 2010 data. The most likely cause of this discrepancy is a small error due to the telescope rotator not having an orientation at exactly $0^{\circ}$ and $90^{\circ}$, but instead being systematically offset by a few degrees. We therefore apply a systematic correction of $+(3.19 \pm 0.73)^{\circ}$ to the measured PA values in $J$, $H$ and $K_{\rm S}$-bands for the 2010 data.

The values of PA in 2013 differed from the optical dust PA by up to $7 \pm 2^{\circ}$. Pereyra \& Acosta-Pulido (in preparation) found that the PA of the polarized standard stars differed from their known optical PA values by up to $\sim 5^{\circ}$ (see their tables 1 and 2), which could be due to the angle of the HWP being offset from the camera angle by this small amount. We therefore add $\pm 5.0^{\circ}$ to the errors of PA for the 2013 data in which the HWP was used. For the 2013 dataset we were also able to measure the FLP of two field stars. This was not possible in 2010 since the camera rotation results in a small, 1' $\times$ 1' field being observed at both rotation angles, whereas the half-wave plate does not rotate the camera and so FLP can be measured in the full 4' $\times$ 1' field. We measured the polarization of two field stars and found them to be polarized, with FLP values a factor of 1.06--1.40 greater than Cyg X--1 in all four filters. The PA of the polarization of the two field stars agreed to within $7 \pm 4^{\circ}$ of the optical interstellar value for Cyg X--1, indicating that the field stars are also polarized due to interstellar dust in the same direction as Cyg X--1. The field stars are fainter, and their polarization errors are larger, than those of Cyg X--1, and the exact optical PA due to dust may differ from star to star.

Using field stars from the Two Micron All Sky Survey \citep[2MASS;][]{skruet06}, we measure magnitudes of $J = 6.909 \pm 0.036$, $H = 6.699 \pm 0.018$ and $K_{\rm S} = 6.572 \pm 0.023$ for Cyg X--1 in 2010, which are on average just 0.05 mag fainter than the 2MASS listed magnitudes for the X-ray binary. In 2013 the magnitudes were $J = 6.975 \pm 0.029$, $H = 6.734 \pm 0.019$ and $K_{\rm S} = 6.617 \pm 0.017$ (we were unable to flux calibrate the $Z$-band data since it is not included in 2MASS); an average of 0.10 mag fainter than the 2MASS magnitudes. Light curves were also produced from each of the individual exposures in 2010 when the conditions were more favourable, and we found that the source is intrinsically variable on short timescales in the IR. We measured the fractional rms to be 4.6 per cent in $J$, 4.3 per cent in $H$ and 4.9 per cent in $K_{\rm S}$-band. The time resolution differed between the filters, so we binned the data such that the time resolution is 16 seconds in all filters, obtaining rms values of $3.2 \pm 1.4$ per cent in $J$, $4.3 \pm 0.6$ per cent in $H$ and $1.5 \pm 0.4$ per cent in $K_{\rm S}$-band (variability was detected at the $2.4 \sigma$, $6.9 \sigma$ and $3.6 \sigma$ confidence levels in the three filters, respectively).

\begin{table*}
\begin{center}
\caption{Multi-wavelength linear polarization measurements of Cyg X--1.}
\vspace{-2mm}
\begin{tabular}{lllllllll}
\hline
Waveband&log($\nu$; Hz)&MJD&X-ray&FLP (\%)&FLP (\%)&PA ($^{\circ}$)&PA ($^{\circ}$)&Ref.\\
&&&state&Observed&Model&Observed&Model&\\
\hline
5 GHz&9.70&49917&hard&$< 10$&9.9&--&69.5&1\\
10.3 $\mu$m (Si-4-band)&13.46&56510, 56570&soft&$< 8.53$&0.00&--&136.8&2\\
2.16 $\mu$m ($K_{\rm S}$-band)&14.14&55365&hard&$0.84 \pm 0.08$&0.79&$142.8 \pm 3.0$&145.8&2\\
2.16 $\mu$m ($K_{\rm S}$-band)&14.14&56548&soft&$0.76 \pm 0.07$&0.51&$131.5 \pm 7.7$&136.8&2\\
1.65 $\mu$m ($H$-band)&14.26&55365&hard&$0.96 \pm 0.06$&1.23&$136.1 \pm 1.8$&139.7&2\\
1.65 $\mu$m ($H$-band)&14.26&56548&soft&$1.11 \pm 0.07$&1.10&$144.0 \pm 6.9$&136.8&2\\
1.25 $\mu$m ($J$-band)&14.38&55365&hard&$1.95 \pm 0.07$&2.07&$137.5 \pm 1.9$&137.7&2\\
1.25 $\mu$m ($J$-band)&14.38&56548&soft&$2.00 \pm 0.07$&2.00&$141.0 \pm 6.1$&136.8&2\\
1.03 $\mu$m ($Z$-band)&14.46&56550&soft&$2.72 \pm 0.07$&2.79&$142.6 \pm 5.8$&136.8&2\\
0.64 $\mu$m ($R$-band)&14.67&47039--44, 47337--43&unknown&$4.40 \pm 0.08$&4.55&$140.8 \pm 0.5$&136.9&3\\
0.55 $\mu$m ($V$-band)&14.74&47039--44, 47337--43&unknown&$4.77 \pm 0.23$&4.78&$141.4 \pm 1.4$&136.9&3\\
0.44 $\mu$m ($B$-band)&14.83&47039--44, 47337--43&unknown&$4.70 \pm 0.30$&4.70&$141.8 \pm 1.8$&136.9&3\\
0.37 $\mu$m ($U$-band)&14.91&42304--16&unknown&$4.35 \pm 0.16$&4.27&$139.7 \pm 1.0$&136.8&4\\
0.40--0.90 $\mu$m$^a$&14.52--14.87&53951, 53955&hard&3.3--4.8&3.4--4.5&135.9--137.1&137.0&5\\
2.6 keV&17.80&42724--43457$^b$&hard&$2.44 \pm 1.07$&3.4&$162 \pm 13$&159.5&6\\
5.2 keV&18.10&42724--43457$^b$&hard&$5.3 \pm 2.5$&3.5&$155 \pm 14$&159.5&6\\
130--230 keV&19.50--19.75&52797--55184$^c$&hard&$< 20$&5--11&--&159.5&7\\
230--370 keV&19.75--19.95&52797--55184$^c$&hard&$41 \pm 9$&11--32&$47 \pm 4$&159.5&7\\
230--850 keV&19.75--20.31&52797--55184$^c$&hard&$76 \pm 15$&11--72&$42 \pm 3$&159.5&7\\
400--2000 keV&19.99--20.68&--$^d$&hard$^d$&$67 \pm 30$&38--80&$40 \pm 10^e$&159.5&8\\
\hline
\end{tabular}
\end{center}
In the fifth and seventh columns the errors on the observed FLP and PA are given at the 1 $\sigma$ level (or for the optical photometric data, they represent the standard deviation of the values taken over all orbital phases). For the model values in columns six and eight, the model for the soft state assumes no jet synchrotron component in the IR.
$^a$Includes spectroscopic data. $^b$Three pointings were used over a three year period; for exact dates see \cite{longet80}. $^c$Nine pointings were used over a six year period; for exact dates see table 1 of \cite{jouret12}. $^d$The data were taken between 2003 and 2009 (no exact dates are given). $^e$The correct value of PA is given in \cite{jouret12}.
References:
(1) \cite{stiret01};
(2) This paper;
(3) \cite{dola92};
(4) \cite{noltet75};
(5) \cite{nagaet09};
(6) \cite{longet80};
(7) \cite{jouret12};
(8) \cite{lauret11}.
\end{table*}

\subsection{Multi-wavelength data collection}

Cyg X--1 is bright, persistent and has been studied extensively for decades. As such, it boasts one of the most well sampled multi-wavelength SEDs of any X-ray binary, spanning $\sim 16$ orders of magnitude in frequency, from the MHz low-frequency radio regime to the GeV high energy $\gamma$-ray regime. Polarimetric measurements have also been made at radio, optical, UV, X-ray and $\gamma$-ray energies. Here, we combine our new mid-IR and NIR data with those previously reported in the literature. Table 4 summarizes the flux measurements of Cyg X--1 collected for this paper, and all polarization measurements of the source are given in Table 5. For many measurements, FLP and its error were calculated by propagating the errors associated with the Stokes parameters $q$ and $u$. We therefore take into account polarization bias \citep{wardkr74} for measurements in which this was not already accounted for. Polarization bias has the effect of increasing the estimated FLP if the errors on $q$ and $u$ are large (usually due to low signal-to-noise ratio; S/N), because FLP is a positive quantity whereas $q$ and $u$ can be positive or negative. The bias-corrected polarization is $FLP_{\rm corr} = \sqrt{1-(\Delta FLP_{\rm obs} / FLP_{\rm obs})^2}$, where $FLP_{\rm obs}$ and $\Delta FLP_{\rm obs}$ are the estimated FLP using the standard formula $FLP = \sqrt{q^2 + u^2}$, and its error by propagating the errors in $q$ and $u$. The reported measurements of FLP generally have high S/N, and $FLP_{\rm obs} - FLP_{\rm corr}$ was found to be $\leq 0.02$ per cent for all optical, UV  and our new NIR data.

Data were collected during periods in which the source was in a hard X-ray state (when this was known), since this is when compact jets are expected to be produced. For all data taken in or after 1996, the X-ray all-sky monitors of the Rossi X-ray Timing Explorer \citep*[RXTE;][]{bradet93} and the Monitor of All-sky X-ray Image \citep[MAXI;][]{matset09} were used to classify the X-ray state of Cyg X--1 on each date, using the classification scheme of \cite{grinet13}. We find that Cyg X--1 was in the hard state when our NIR data were taken in 2010. However, in 2013, Cyg X--1 resided in the soft state on all dates when our mid-IR and NIR data were acquired. We therefore caution that the compact jet may not make a contribution to the polarization in the 2013 data since the source was not in the hard state. We include UV flux spectra that were taken during a soft state, but since the unvarying companion dominates at these wavelengths (the corona and jet are $> 2$--3 orders of magnitude fainter), we can include these flux measurements in our SED. Absorbed data were de-reddened using the extinction to the source, $A_{\rm V} = 2.95$ mag \citep*{wuet82,rahoet11,xianet11} and adopting the IR/optical/UV extinction laws of \cite*{cardet89}, \cite{pei92} and \cite{chiati06}. Unabsorbed X-ray and mid-IR spectra were taken from \cite{rahoet11} (observation 1 as defined by the authors, during which the source was in a hard state), and unabsorbed X-ray and $\gamma$-ray data were taken from \cite{zdziet12} and \cite{malyet13}.

\begin{figure*}
\centering
\includegraphics[width=8.3cm,angle=0]{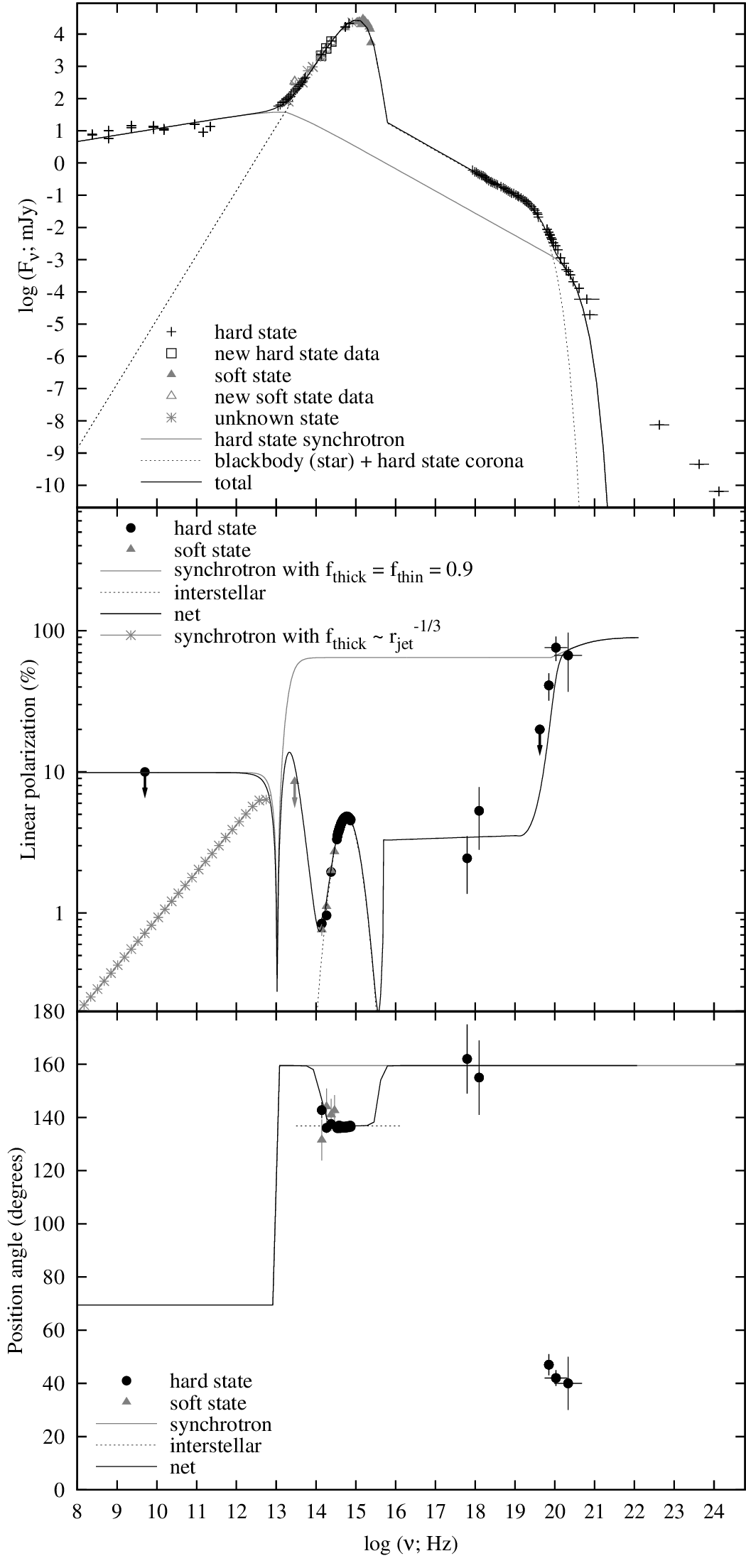}
\includegraphics[width=8.3cm,angle=0]{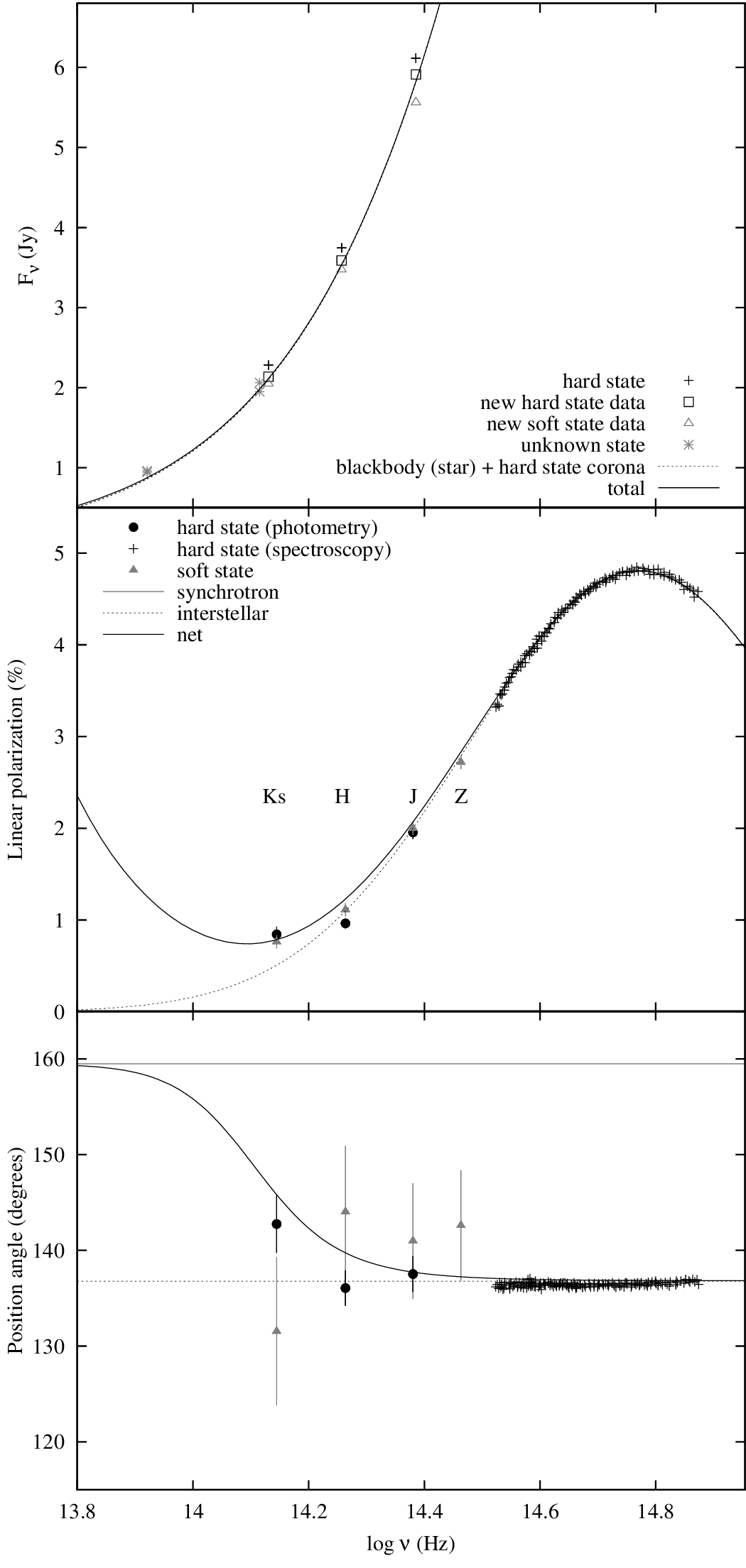}
\caption{\emph{Left:} Radio to $\gamma$-ray flux density spectrum ($F_{\nu}$; upper panels), FLP spectrum (centre panels) and polarization PA (lower panels) of Cyg X--1. \emph{Right:} The same as the left panels but just the NIR--optical region, showing how the interstellar model approximates the IR--optical data \citep[see also][]{nagaet09}, except in $K_{\rm S}$-band most evidently in the hard state. See Tables 4 and 5 for data references and Table 6 for our model parameters.}
\end{figure*}

\section{A toy model for the multi-wavelength flux and polarization of Cyg X--1}

\subsection{The flux spectrum}

The broadband, radio to $\gamma$-ray flux density ($F_{\nu}$) spectrum of Cyg X--1 in the hard state is presented in the upper left panel of Fig. 1, and the same is presented in Fig. 2 as a SED ($\nu F_{\nu}$). We also show UV and IR soft state data as grey triangles. The supergiant O star dominates the IR/optical/UV emission and can be approximated by a single temperature blackbody. While the jet produces the flat/inverted optically thick radio synchrotron spectrum, \cite{fendet00} showed that this spectrum extends to millimetre wavelengths, and \cite{rahoet11} found evidence for variable synchrotron emission at mid-IR wavelengths. Broadband models applied to data of Cyg X--1 \citep*[e.g.][]{market05,nowaet11,rahoet11,zdziet12,zdziet13} typically include a Comptonized corona which most likely dominates the X-ray flux, and a jet, which dominates the radio--mm regime and may make a significant contribution to the IR, X-ray and $\gamma$-ray flux.

Here, we adopt a simple, toy model of a synchrotron jet, a blackbody from the companion star, and a Comptonized corona approximated by a power law with an exponential cutoff. The aim is to approximately reproduce the observed spectrum phenomenologically, in order to use this as an input spectrum for the polarization model described below. The jet consists of a broken power law describing the optically thick and optically thin regions of the synchrotron spectrum, with some curvature at the frequency of the spectral break, $\nu_{\rm b}$ \citep[e.g.][]{blanko79}. \cite{rahoet11} performed spectral modelling of broadband SEDs of Cyg X--1 which included mid-IR Spitzer spectra, and their fits favoured a break in the jet spectrum at $\nu_{\rm b} \sim 3 \times 10^{13}$ Hz, with the optically thin power law extending to higher frequencies. In order to explain the high energy tail in the $\gamma$-ray spectrum and the high level of polarization, it was claimed that this optically thin synchrotron power law extends to the $\gamma$-ray regime \citep[e.g.][]{lauret11,rahoet11,jouret12}. This has recently been tested by applying jet models to the broadband SED of Cyg X--1 in the hard state \citep{zdziet12,malyet13,zdziet13}, including constraints from new $\gamma$-ray detections at GeV energies by FERMI \citep[both the average hard state flux and flares have been detected;][]{malyet13,bodaet13}. The flux spectrum is consistent with the optically thin synchrotron power law from the jet to extend to $\gamma$-ray energies and account for the MeV tail with a cut-off in that regime, and this is the preferred interpretation if the high polarization measurements are robust \citep{zdziet12,zdziet13,malyet13}.

These results have direct implications for the expected polarization as a function of frequency in the SED of Cyg X--1, since the polarization properties of optically thin synchrotron emission are well understood, allowing us to use polarimetry to test these models. Below we compare the observed broadband polarization properties of Cyg X--1 with that expected from a synchrotron jet, the spectrum of which (and its contribution to the total flux) is defined here by the model approximating the flux spectrum.

Initially, we adopted the best fit model parameters reported by \cite{rahoet11} in order to explain the jet spectrum. We find that the model can approximately reproduce the radio--mm data \citep[we include significantly more data covering a larger frequency range compared to][ but our data are not quasi-simultaneous]{rahoet11}; the scatter in the data reflect the variability seen of a factor of a few in flux in the hard state due to the data not being simultaneous. However, the model underpredicts the mid-IR flux at $\sim 10^{13}$ Hz. \cite{rahoet11} account for this by including an additional power law, which is claimed to be bremsstrahlung from the stellar wind of the companion. We adopt a more inverted optically thick spectral index ($\alpha_{\rm thick} = 0.2$) and find that a bremsstrahlung component is no longer required. The radio to UV data and the high energy tail can be approximated by the jet component and the blackbody from the companion. Although the radio to mm SED of Cyg X--1 was shown to be approximately flat \citep{fendet00}, the average radio spectral index that included contemporaneous low-frequency radio data was found to be $\alpha_{\rm thick} = 0.26 \pm 0.10$ \citep[at 0.6--15 GHz;][]{pandet07}. We therefore favour a model without a bremsstrahlung component and with an inverted optically thick jet spectrum. Our model is presented in the upper panels of Fig. 1.

The broken power law synchrotron spectrum with an exponential cutoff at high energies is described by,
\begin{eqnarray}
F_{\rm syncBPL} / \rm mJy = \left\{
\begin{array}{c l}     
    n_1 \nu^{\alpha_{\rm thick}} & : \nu < \nu_{\rm b}\\
    \\
    n_2 \nu^{\alpha_{\rm thin}} & : \nu_{\rm b} \leq \nu \leq \nu_{\rm cut}\\
    \\
    n_3 e^{-n_4 \nu / \nu_{\rm cut}} & : \nu > \nu_{\rm cut},
\end{array}\right.
\end{eqnarray}
where $n_{\rm i}$ are normalization constants. The optically thin spectral index is defined by the electron energy distribution, $\alpha_{\rm thin} = 0.5 (1-p)$. 
We then include a term to introduce curvature between the optically thick and optically thin power laws,
\begin{eqnarray}
n_{\rm curve} = \left\{
\begin{array}{c l}     
    ((n_5 n_1 \nu^{\alpha_{\rm thick}}) / (n_2 \nu^{\alpha_{\rm thin}})) + 1 & : \nu < \nu_{\rm b}\\
    \\
    ((n_5 n_2 \nu^{\alpha_{\rm thin}}) / (n_1 \nu^{\alpha_{\rm thick}})) + 1 & : \nu \geq \nu_{\rm b}.
\end{array}\right.
\end{eqnarray}
The final curved synchrotron spectrum is,
\begin{eqnarray}
F_{\rm sync} / \rm mJy = F_{\rm syncBPL} / n_{\rm curve}.
\end{eqnarray}
In synchrotron jet models, several parameters can vary the amount of curvature, such as the magnetic field profile and deviations in the relativistic particle density close to the base of the jet. We therefore do not calculate the curvature because these parameters cannot be measured, but instead choose a value for the curvature constant $n_5$ based on the the amount of curvature that well describes the jet break of GX 339--4, in which the curvature is clearly visible \citep{gandet11}.

A simple blackbody (Planck's law) describes the companion star,
\begin{eqnarray}
F_{\rm BB} / \rm mJy = n_6 \nu^3 / (e^{(h \nu)/(k_{\rm B}T)} -1),
\end{eqnarray}
where $h$ is the Planck constant, $k_{\rm B}$ is the Boltzmann constant and $T$ is the blackbody temperature. The Comptonized corona is approximated by a power law with a high energy exponential cutoff at $\nu_{\rm compcut}$,
\begin{eqnarray}
F_{\rm comp} / \rm mJy = \left\{
\begin{array}{c l}     
    n_7 \nu^{\alpha_{\rm comp}} & : \nu \leq \nu_{\rm compcut}\\
    \\
    n_8 e^{-n_9 \nu / \nu_{\rm compcut}} & : \nu > \nu_{\rm compcut}.
\end{array}\right.
\end{eqnarray}
We consider a low energy limit to this power law at $\nu = 5 \times 10^{15}$ Hz, below which the companion dominates the flux anyway. Finally, the total spectrum is the sum of the synchrotron, blackbody and Comptonized corona,
\begin{eqnarray}
F_{\rm total} / \rm mJy = F_{\rm sync} + F_{\rm BB} + F_{\rm comp}.
\end{eqnarray}

We see from Fig. 1 that as expected, the jet dominates the hard state radio--mm--mid-IR spectrum, the companion becomes the brightest emitter in the mid-IR--NIR--optical--UV regime, the corona dominates the X-ray flux and the jet spectrum accounts for the $\gamma$-ray MeV tail \citep*[at $\nu \sim 10^{20}$ Hz; see also][]{malzet09,lauret11,rahoet11,jouret12}. It has been demonstrated that synchrotron self-Compton emission (SSC) and/or Compton upscattering of blackbody stellar emission can account for the GeV ($\nu \sim 10^{22}$--$10^{24}$ Hz) flux in the hard state \citep{malyet13,zdziet13}. Since no GeV polarization measurements have been made, and the GeV-emitting component seems not to dominate at energies lower than the GeV regime, we have no need to include an extra component in our model at GeV energies.

\begin{figure*}
\centering
\includegraphics[width=16.0cm,angle=0]{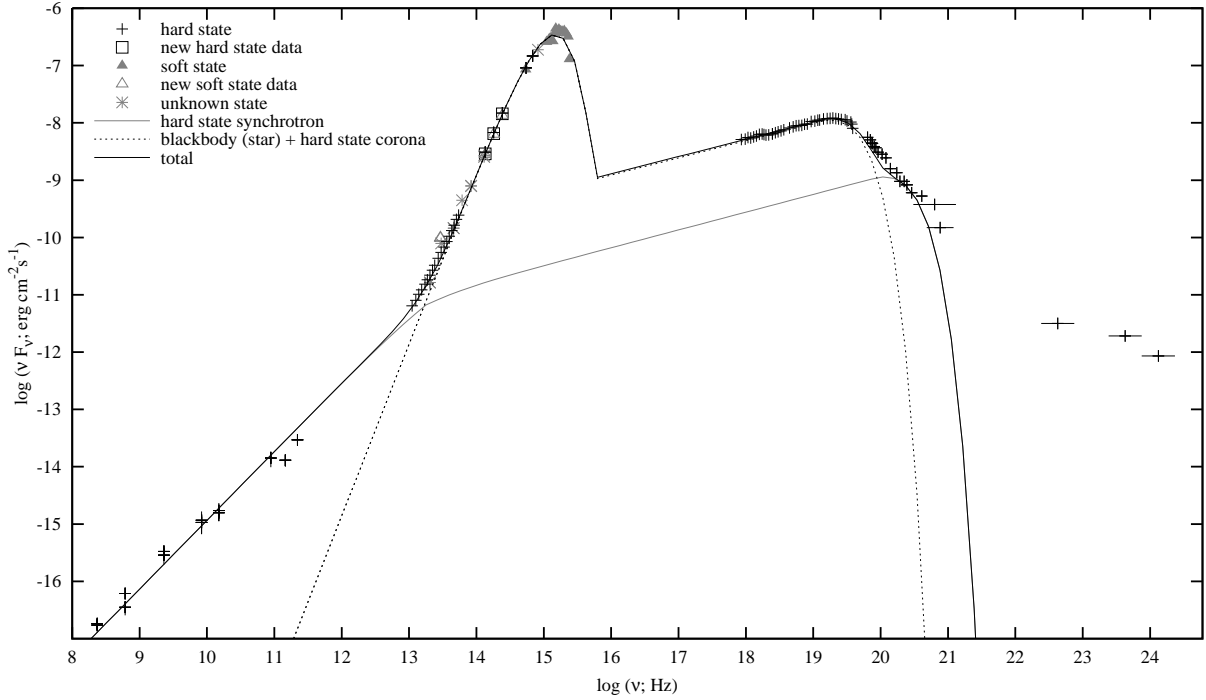}
\caption{Radio to $\gamma$-ray spectral energy distribution ($\nu F_{\nu}$) of Cyg X--1. See Table 4 for data references and Table 6 for our model parameters.}
\end{figure*}

\subsection{The polarization spectrum}

\cite{west59} first treated the linear polarization properties of optically thin synchrotron emission from electrons in a uniform magnetic field. Since then several works have developed and extended the calculations to include power-law electron energy distributions and nonuniform magnetic fields \citep[e.g.][]{ginzsy65,nord76}. The general case for optically thin synchrotron emission from an electron population with an arbitrary distribution of energies and an arbitrary magnetic field configuration was derived by \cite{bjorbl82}. The expected linear polarization here is \citep[see also][]{rybili79},
\begin{eqnarray}
  FLP_{\rm thin} = f \frac{p + 1}{p + 7/3} = f \frac{1 - \alpha_{\rm thin}}{5/3 - \alpha_{\rm thin}},
\end{eqnarray}
where $f$ represents the ordering of the magnetic field and takes values between zero (nonuniform, no net field orientation) and one (a perfectly uniform, aligned field), and $p$ is the electron energy distribution. For optically thin synchrotron emission with a typical spectral index of $\alpha_{\rm thin} \sim -0.7$ ($p = 2.4$), the maximum polarization (for an ordered magnetic field, i.e. $f = 1$) is therefore $FLP_{\rm thin} = 72$ per cent. If the spectral index is steeper the polarization can be higher still, with $FLP_{\rm thin} = 82$ per cent for $\alpha_{\rm thin} = -2$.

For a homogeneous synchrotron source with a power law distribution of electron energies, the FLP is frequency-independent, but curvature in the spectrum of electron energies can result in frequency-dependent FLP \citep{bjor85}. In blazars, the superposition of ordered plus chaotic magnetic field components results in curvature in the SED, and frequency dependency of polarization \citep[see also, e.g.][]{valtet91,barret10}. In BHXBs, the optically thin spectrum from the compact jet is consistent with a single power law when it is measured well \citep[e.g.][]{hyneet03,russet13b}, with minimal curvature, which likely originates in a single population of electrons with a power law energy distribution. There is curvature at the high energy cutoff however, so we expect a change in the FLP at the highest energies. Above the high energy cutoff frequency,
\begin{eqnarray}
  FLP_{\rm cut} = f \frac{1 - \alpha_{\rm cut}}{5/3 - \alpha_{\rm cut}},
\end{eqnarray}
where $\alpha_{\rm cut}$ is the spectral index defined by $F_{\rm syncBPL}$. Optically thick (absorbed) synchrotron radiation has a flux spectrum $F_{\nu} \propto \nu^{5/2}$ for a single electron distribution, and is expected to be less polarized than optically thin synchrotron \citep[e.g.][]{blanet02},

\begin{eqnarray}
  FLP_{\rm thick} = f \frac{3}{6 p + 13},
\end{eqnarray}
with a position angle that differs by $90^{\circ}$ to that of optically thin synchrotron polarization \citep[see also][]{alle70,joneod77,rudnet78}. The maximum polarization from this spectrum is $FLP_{\rm thick} = 11$ per cent for $p = 2.4$ ($\alpha_{\rm thin} = -0.7$). If the level of ordering of the magnetic field remains constant down the length of the jet, with a constant PA, then equation (9) could describe the polarization expected from a flat/inverted optically thick jet spectrum composed of overlapping synchrotron spectra. If however, the ordering changes or the PA varies with distance from the jet base (e.g. in helical fields), the polarization properties will change as a function of frequency in the optically thick spectrum. We apply equation (9) to the data of Cyg X--1 so that we can predict the FLP and PA of the optically thick spectrum in the case of a constant ordering and PA of the magnetic field along the jet. We also consider a jet in which the field ordering changes with distance along the jet. The emitting region in standard flat spectrum jets is located at a distance from the jet base approximately inversely proportional to the frequency of the emission, $r_{\rm jet} \sim \nu^{-1}$ \citep{blanko79}. Therefore, if the magnetic field is increasingly tangled down the length of the jet such that $f \propto r_{\rm jet}^{-1}$ for example, the polarization would decrease rapidly from the jet break to lower frequencies and would be negligible at in the mm--radio regime. We consider different values of $\beta$, where,
\begin{eqnarray}
  f \propto r_{\rm jet}^{-\beta} \propto \nu^{\beta} & : \nu < \nu_{\rm b},
\end{eqnarray}
in the optically thick synchrotron spectrum. Here, $\beta$ is the index representing the dependency of the ordering of the magnetic field on the distance along the jet, with $\beta = 0$ indicating a constant field ordering. Observationally comparing $FLP_{\rm thin}$ with the frequency-dependent $FLP_{\rm thick}$ therefore probes this dependency of field ordering on distance. The maximum $FLP_{\rm thick}$ of $\simlt 11$ per cent should hold in all cases, and radio data of several sources have indicated that $FLP_{\rm thick}$ is typically a few per cent, with up to $\sim 5$--8 per cent reported in a few sources \citep[e.g.][]{gallet04,brocet13,curret13}.

The above equations are used to predict the FLP of the jet spectrum. At IR/optical/UV frequencies the companion star dominates which, by definition for a blackbody, is expected to be unpolarized. The observed FLP will therefore drop at IR and higher frequencies as the synchrotron contribution to the total flux decreases with increasing frequency. However, optical FLP values of 3--5 per cent are well documented in the literature for Cyg X--1, and almost all of this FLP has an interstellar dust origin. A small fraction of optical FLP has been found to be due to scattering in the stellar wind of the companion and varies on the orbital period, and an additional long-term (decades) variation which may be caused by scattering by a varying asymmetric stellar wind or other circumstellar matter \citep[for a detailed study see][]{nagaet09}. We adopt the interstellar dust polarization model of \cite*{serket75}, which has been fitted to optical spectra of Cyg X--1 by \cite{nagaet09}, and we use the parameters measured by \cite{nagaet09}. The detail of the model and data in the optical--IR region of the spectrum can be seen in right panels of Fig. 1.

Photons from the Comptonized corona may be emitted isotropically and if no relativistic, beaming or bulk motion effects are present then one may expect a very low net polarization for the corona. Comptonization of disc photons is expected to be low \citep{schnkr09,maitpa11} since the disc photons have a low net polarization \citep[even when taking relativistic effects into account, the peak FLP is predicted to be 1--5 per cent; e.g.][]{dextqu12}. Relitivistic reflection on the disc surface could produce up to $\sim 10$ per cent observed X-ray polarization depending on the height of the primary source and the viewing angle \citep[e.g.][]{dovcet11,goosma11,mariet13}. Compton scattering of unpolarized disc photons by a relativistic jet can produce FLP $\sim 3$--20 per cent depending on the viewing angle \citep*{mcnaet09}. In the commonly used Comptonized corona model the seed photons are from the disc (but some may also be from the jet), so the corona is expected to be not much more than 5 per cent polarized \citep[see also][]{pout94,schnkr10,veleet13}. In our toy model we assume zero polarization for this component, $FLP_{\rm comp} = 0$.

SSC emission can be polarized by up to $\sim 30$--50 per cent of the original synchrotron source polarization, for Lorentz factors of 2--10 (likely typical for BHXBs) \citep{celoma94,pout94,mcnaet09,kraw12}. For a maximum synchrotron polarization of $\sim 82$ per cent, the maximum FLP of the SSC component is $\sim 25$--41 per cent, which is significantly lower than the highest $\gamma$-ray FLP of $76 \pm 15$ per cent detected from Cyg X--1. Unlike synchrotron emission, FLP of SSC emission is dependent on frequency and viewing angle. Broadband models predict a more curved spectrum for the SSC component \citep[e.g.][]{market05,nowaet11}, and recent works favour this component to peak in the GeV regime \citep{malyet13,zdziet13}. For a recent discussion on the different sources of X-ray polarization in BHXBs, see \cite{schnet13}.

Since different components in our model produce emission with different polarization position angles, it is necessary to calculate the Stokes parameters $q$ and $u$ for each component at each frequency from the known FLP and PA values, using the standard equations $FLP = \sqrt{q^2 + u^2}$ and $PA = 0.5 tan^{-1}(u/q)$. The positive/negative signs of $q$ and $u$ are lost due to the square root, so we multiple by $1$ or $-1$ depending on the value of PA in order to derive a consistent solution. For the interstellar dust law, the PA has been measured from the data and is constant with frequency \citep{nagaet09}. For synchrotron we treat PA as an input free parameter, with $PA_{\rm thick}$ and $PA_{\rm thin}$ differing by $90^{\circ}$. The values of $q$ are thus calculated,
\begin{eqnarray}
  q_{\rm thick} = \frac{FLP_{\rm thick}}{(tan^2(2(PA_{\rm thin}-90^{\circ})) + 1) ^{1/2}} \\
  q_{\rm thin} = \frac{FLP_{\rm thin}}{(tan^2(2 PA_{\rm thin}) + 1) ^{1/2}} \\
  q_{\rm cut} = \frac{FLP_{\rm cut}}{(tan^2(2 PA_{\rm thin}) + 1) ^{1/2}} \\
  q_{\rm dust} = \frac{FLP_{\rm dust}}{(tan^2(2 PA_{\rm dust}) + 1) ^{1/2}} \\
  q_{\rm comp} = 0.
\end{eqnarray}
The values of $u$ are simply $u = \sqrt{FLP^2 - q^2}$ for each component. The whole polarization spectrum for synchrotron is $q_{\rm sync}$ and includes a smooth transition between $q_{\rm thick}$ and $q_{\rm thin}$ around $\nu_{\rm b}$ that corresponds to the curved flux spectrum. The net polarization is then calculated in $q$, $u$ space,
\begin{eqnarray}
  q_{\rm total} = q_{\rm sync} \frac{F_{\rm sync}}{F_{\rm total}} + q_{\rm dust} \\
  u_{\rm total} = u_{\rm sync} \frac{F_{\rm sync}}{F_{\rm total}} + u_{\rm dust}.
\end{eqnarray}

\begin{table}
\begin{center}
\caption{Model parameter values.}
\vspace{-2mm}
\begin{tabular}{lll}
\hline
&Parameter&Value\\
\hline
\multicolumn{2}{l}{\emph{Synchrotron jet:}}&\\
&$\nu_{\rm b}$		&$1.3 \times 10^{13}$ Hz (23 $\mu$m) \\
&$\nu_{\rm cut}$		&$8 \times 10^{19}$ Hz (330 keV) \\
&$p$				&2.38 \\
&$\alpha_{\rm thin}$	&$-0.69$ \\
&$\alpha_{\rm thick}$	&$+0.20$ \\
&$f$				&0.9 \\
&$PA_{\rm thin}$		&$159.5^{\circ}$ \\
&$PA_{\rm thick}$	&$69.5^{\circ}$ \\
&$PA_{\rm jet}$	&$159.5^{\circ}$ \\
\multicolumn{2}{l}{\emph{O star companion:}}&\\
&$T$				&$1.8 \times 10^{4}$ K \\
\multicolumn{2}{l}{\emph{Interstellar dust:}}&\\
&$FLP_{\rm dust,max}$	&$4.8\%$ \\
&$\nu_{\rm dust,max}$	&$5.9 \times 10^{14}$ Hz (0.5 $\mu$m) \\
&$PA_{\rm dust}$		&$136.8^{\circ}$ \\
\multicolumn{2}{l}{\emph{Comptonized corona:}}&\\
&$\nu_{\rm compcut}$	&$1.1 \times 10^{19}$ Hz (45 keV) \\
&$\alpha_{\rm comp}$	&$-0.70$ \\
&$FLP_{\rm comp}$	&0 \\
\hline
\end{tabular}
\end{center}
\end{table}

\subsection{Chosen model parameters}

The resulting FLP spectrum and PA spectrum are shown in the centre and lower panels of Fig. 1, respectively. We adopt the values $f = 0.9$ and $PA_{\rm thin} = 159.5^{\circ}$ \citep[which is the mean position angle of the radio jet on the plane of the sky;][]{stiret01}. By choosing these values the observed FLP and PA at all frequencies can be recovered by the model, with one exception: the PA of the $\gamma$-ray polarization differs to that of the model by $\sim 60^{\circ}$. The high value of $f$ is required in order to produce a synchrotron spectrum that can account for the very high $\gamma$-ray FLP. We find that by adopting this value of $f$, the expected lower FLP in X-ray is also consistent with the model, since the (assumed to be unpolarized) corona dominates the X-ray flux, with the synchrotron power law contributing $\sim 5$ per cent of the flux. In the case of a constant value of $f$ along the length of the jet ($\beta = 0$), the model predicts that the FLP in radio is $\sim 10$ per cent, which is just consistent with the observed upper limit \citep{stiret01}. However, if the magnetic field becomes less ordered with distance along the jet such that $f \propto r_{\rm jet}^{-1/3}$ ($\beta = 1/3$), the predicted radio polarization is $< 1$ per cent. For many BHXBs the flat spectrum radio emission is polarized at a level of a few per cent \citep[e.g.][]{gallet04,brocet13}, so for Cyg X--1 the likely value of $\beta$ is between 0 and 1/3.

The optical and NIR FLP in the hard state can be well described by the model; we find that the $J$ and $H$-band data fit very well on the extrapolation of the interstellar dust model, but the FLP is slightly higher than expected from the dust model in the lowest frequency filter, $K_{\rm S}$-band (see Fig. 1, right panels). The model predicts an upturn in the FLP at frequencies lower than $\sim K_{\rm S}$-band, whereby the highly polarized jet synchrotron emission starts to make a stronger contribution. At the frequency of the jet spectral break, the FLP is expected to drop to zero and at lower frequencies the FLP becomes that expected for optically thick synchrotron emission. The deviation of the $K_{\rm S}$-band data from the interstellar model is most prominent in the hard state data from 2010, but the 2013 FLP taken in the soft state also appear to be higher than the interstellar model, possibly implying a synchrotron component present in the soft state. However, the measured mid-IR polarization of $FLP = 0.82 \pm 2.57$ per cent from the 2013 soft state data is significantly lower than what the model predicts for the hard state jet at this frequency. The $3 \sigma$ upper limit of $FLP < 8.53$ per cent is still slightly lower than the model prediction of about 11 per cent for the hard state. In addition, the NIR fluxes in 2013 were significantly fainter (by $3.6 \sigma$, $4.3 \sigma$ and $6.8 \sigma$ confidence levels in $J$, $H$ and $K_{\rm S}$) than the 2MASS fluxes, whereas the 2010 fluxes are all consistent with 2MASS within $3 \sigma$. This is consistent with a flux drop due to the lack of jet contribution in the soft state. A lower mid-IR flux in the soft state compared to the hard state was also seen by \cite{rahoet11}.

By adopting the value $PA_{\rm thin} = 159.5^{\circ}$, the model is also able to recover the observed X-ray PA. We find that the $\gamma$-ray FLP actually requires the X-rays to be polarized on a level of a few per cent, as observed, under the assumption of a synchrotron jet origin. The optical--NIR FLP and PA values are consistent with interstellar dust, but we find that around the NIR $K_{\rm S}$-band, the model predicts a smooth shift in PA between that expected from the optically thin synchrotron jet ($PA_{\rm thin} = 159.5^{\circ}$) and that of the interstellar dust component ($PA_{\rm dust} = 136.8^{\circ}$). The observed $K_{\rm S}$-band PA in the hard state is $\sim 6^{\circ}$ higher than the $J$ and $H$-bands, and consistent with this smooth transition. Since the FLP is also higher than expected from interstellar dust, the results imply that the polarimetric signature of the jet is detected at $2 \mu$m. In the soft state, the PA values have larger errors due to the use of the HWP. In $Z$, $J$ and $H$-bands, the PA is consistent with the models both with and without synchrotron. In $K_{\rm S}$-band the PA is consistent within $1 \sigma$ with the interstellar model but not the model with synchrotron included, which suggests that the synchrotron jet does indeed make a contribution in the hard state, but not in the soft state, as expected. The only NIR PA value that is inconsistent with the interstellar value is the $K_{\rm S}$-band PA in the hard state. It is worth noting that a weak radio jet may exist in the soft state of Cyg X--1 \citep{rushet12} but is much fainter than the hard state jet \citep[the 15 GHz radio flux density in the soft state can be up to two orders of magnitude lower than the average hard state; e.g.][]{zdziet11}. In other BHXBs the soft state jet, if it exists, is hundreds of times fainter in radio than the hard state jet \citep{russet11a,coriet11}.

The model is able to reproduce the very high observed $\gamma$-ray FLP, and a slight increase in FLP with energy is expected around the high-energy cut-off where the synchrotron spectral index becomes steeper. The model cannot explain the PA of the highly polarized $\gamma$-ray emission, which appears to imply a field that is mis-aligned with the jet axis in the $\gamma$-ray emitting region of the jet. In Section 4.1 we discuss several reasons why this could be the case. The model implies a very ordered and very stable magnetic field near the base of the jet of Cyg X--1. The electric vector is parallel to the known radio jet axis $PA_{\rm jet}$, so the magnetic field lines are orientation perpendicular to the jet axis. The model parameter values are given in Table 6.

\section{Discussion}

\subsection{The highly ordered magnetic field in the jet of Cyg X--1}

The results imply that the magnetic field near the base of the jet of Cyg X--1 is highly ordered, orthogonal to the jet axis and stable over several years, since the INTEGRAL polarization is measured over this timescale. This is astonishing because in other X-ray binaries and AGN the magnetic field ordering in this region of the jet is usually found to be much lower. In BHXBs and some neutron star X-ray binaries, the FLP is a few per cent and the magnetic field lines are usually preferentially parallel to the jet axis, with some evidence for variability on short timescales, even in cases where synchrotron emission appears to dominate the flux \citep[see][ and references therein]{russet11b}. However, only a few systems have been studied to date.

Similar values of FLP are found from optically thin synchrotron emission in the compact jets of AGN. Gigahertz peaked-spectrum (GPS) sources and compact steep-spectrum (CSS) sources have jet spectral breaks at radio frequencies, and polarization measurements of the core, optically thin synchrotron emission have been obtained. The polarization is measured to be $\sim 1$--7 per cent in the optically thin regime \citep[for a review, see][]{odea98}. This suggests a similarly tangled magnetic field in the core jet in both AGN and X-ray binaries (but not Cyg X--1). Blazars are synchrotron-dominated sources in which higher levels of FLP have been measured (tens of per cent and variable) but in these AGN, the flux and polarization are boosted by beaming effects \citep[e.g.][]{mars06}. Knots and interactions downstream in AGN jets can also be highly polarized \citep[e.g.][]{saiksa88,listho05,perlet06}; these are analogous to the tens of per cent radio FLP measured from discrete ejecta detached from the core in X-ray binaries \citep[e.g.][]{fendet99,hannet00,brocet07,brocet13}. The flat-spectrum (optically thick) radio jets of AGN are weakly polarized (FLP $\sim 1$--5 per cent in most cases), similar to the optically thick radio jets of X-ray binaries. The magnetic field lines in flat-spectrum AGN cores do not have a preferential orientation, but span a wide range of position angles relative to the direction the jet is travelling in \citep[][]{listho05,helmet07}.

It is unclear why the jet in Cyg X--1 would have a less tangled magnetic field compared to other BHXBs, and AGN. The magnetic field strength, which can be estimated from the optically thick--thin spectral break \citep[e.g.][]{chatet11,gandet11}, is actually similar to other BHXBs \citep{rahoet11,russet13a}, but nevertheless the field appears to be highly ordered compared to other systems. Cyg X--1 is the first high-mass X-ray binary (HMXB) in which the polarimetric signature of the jet has been measured \citep[intrinsic IR polarization was detected in Cyg X--3, but the jet contribution was uncertain;][]{joneet94}. Accretion onto the BH in Cyg X--1 is relatively stable due to the stellar-wind of the companion and the source always accretes at close to $\sim 1$ per cent of the Eddington luminosity. This is in contrast to most BHXBs, which are transient low-mass X-ray binaries (LMXBs) with much weaker stellar winds. It is possible that the stability of the mass accretion rate in Cyg X--1 has caused the magnetic field structure in the inner jet to reach an equilibrium, whereas the rapidly changing accretion rate in transient BHXBs would not allow this to occur, possibly leading to more chaotic field structures in transient BHXBs.

On large scales, HMXB jets (and the photons they produce) interact with the stellar wind of the companion \citep[e.g.][]{fermet09,tavaet09,perubo12,corbet12,zdziet13}, but on the small scales of the inner regions of the jets (considered to be distances $\sim 100 R_{\rm g}$ from the BH; where the gravitational radius is $R_{\rm g} = GM / c^2$) it is less clear why HMXB and LMXB jets would differ. It is worth noting that measurements of the spin of the BH in Cyg X--1 favour a high spin \citep[close to maximally spinning;][]{gouet11,fabiet12}, which may or may not lead to changes in the jet properties \citep*{fendet10,naramc12,steiet13,russet13c}. GRS 1915+105, which is also claimed to have a high BH spin \citep{mcclet06,blumet09}, is not highly polarized at IR wavelengths \citep*{samset96,shahet08}. Whatever the reason is for a more highly ordered field in the jet of Cyg X--1, the result hints at a fundamental difference between the conditions in the inner regions of the jets of HMXBs and LMXBs, and possibly a different jet launching process.

An alternative explanation for the high $\gamma$-ray FLP is that the emitting region is much smaller than the IR and X-ray emitting regions of the jet, possibly indicating a more highly ordered field on smaller spatial scales in the jet. Although the MeV photons are expected to come from the same distribution of electron energies as the X-ray and IR photons in the synchrotron plasma, the cooling time for $\gamma$-ray photons is shorter and so the $\gamma$-ray-emitting electrons radiatively lose their energy
on shorter timescales than X-ray-emitting electrons. The $\gamma$-ray emitting regions would therefore be smaller than the X-ray emitting regions, since the higher energy photons have less time to travel away from their emission sites. Any misalignment in the field orientation between these small regions in the jet will reduce the net polarization measured on large scales, so it is likely that the magnetic field will appear more uniform on small scales, and higher FLP will be produced at $\gamma$-ray energies. If this effect is responsible for the high $\gamma$-ray FLP then the net field ordering could be much lower than $f = 0.9$ and our model could be overestimating the FLP from synchrotron at lower energies than $\gamma$-ray. However, if that is the case then the X-ray and NIR ($K_{\rm S}$-band) FLP values cannot be explained.

The only parameter that our model cannot reproduce is the $\gamma$-ray PA; the observed PA differs by $\sim 60^{\circ}$ to the model prediction. In our simple toy model the PA is constant for optically thin synchrotron emission \citep[see][]{bjorbl82,blanet02}. However, electron distributions with a sharp break or cut-off could produce synchrotron emission with frequency-dependent PA \citep{nord76,bjor85}. In the $\gamma$-ray regime our model favours a curved spectrum, with $\alpha$ decreasing with energy around/above the high-energy cut-off in the synchrotron spectrum. Numerical solutions have shown that the rate of change of PA with frequency can be large, especially for steeper spectra, while FLP may increase by only a factor $\leq 2$ \citep{bjor85}. Since these are numerical and not analytical solutions, we have not included them in our model. Nevertheless, the curved $\gamma$-ray spectrum leads to the possibility that a change of PA could be expected. This shift in PA should only occur in the curved region, not in the power law at X-ray to IR frequencies. In the numerical results of \cite{bjor85}, the PA changes smoothly by up to $\sim 60^{\circ}$ (depending on various parameters) over two orders of magnitude in frequency for a spectrum that curves downwards at higher energies. This may therefore explain the apparent shift of $60^{\circ}$ between the X-ray PA and the $\gamma$-ray PA, both of which originate in the optically thin synchrotron emission from the jet in our model.

Alternatively, the $\gamma$-ray PA could be probing a small region in the jet that has a different field orientation to the larger-scale IR to X-ray emitting region. The jet in the AGN M87 has been spatially resolved at radio frequencies down to a few Schwarzschild radii from the BH, and the opening angle of the jet is seen to collimate, from an $\sim 60^{\circ}$ at $\sim 15$--$50 R_{\rm g}$ to $\sim 5$--$10^{\circ}$ at distances three orders of magnitude larger \citep*[e.g.][]{junoet99,doelet12}. Under the assumption that the jet of Cyg X--1 could also have an opening angle of $60^{\circ}$, if the emission from the jet could be limb-brightened in a small region close to the jet base, the $\gamma$-ray PA could imply a magnetic field that is parallel to the ridge of the jet. This is illustrated by the schematic in Fig. 3. In this scenario, the magnetic field is orthogonal to the direction of the motion of the jet in the IR to X-ray emitting region (as implied by the model) and a small region closer to the BH produces the $\gamma$-ray emission in which the PA is changed by $60^{\circ}$. If the opening angle of the jet in the $\gamma$-ray emitting region is $60^{\circ}$, the magnetic field is parallel to the ridge of the jet. This scenario could account for the apparent shift in PA by $60^{\circ}$ between the X-ray and $\gamma$-ray. However, it is unclear how one small region of the limb-brightened jet would dominate the $\gamma$-ray emission. One may expect that if the jet is limb-brightened then emission should be seen from both `sides' of the jet. If the magnetic field is parallel to the ridge of the jet then an opening angle of $60^{\circ}$ would produce an average, integrated magnetic field that is aligned with the jet axis and would produce lower net FLP. This scenario also requires the $\gamma$-ray emission to originate in a region of the jet that is closer to the BH than the IR to X-ray emitting region. This is not expected since in our model the same distribution of electrons produces the IR to $\gamma$-ray emission. We therefore favour the previous explanation -- the $60^{\circ}$ difference in PA between the X-ray and $\gamma$-ray is likely to be due to the steepening of the spectrum.

\subsection{Evidence for a jet contribution to the X-ray and $\gamma$-ray flux}

The models here generally assume the optically thin synchrotron power law from the jet extends to X-ray energies, providing a low but significant contribution to the X-ray luminosity. In Cyg X--1, this power law dominates the $\gamma$-ray luminosity. Generally, the origin of the X-ray emission in X-ray binaries is still discussed at length, and the jet contribution is currently a hot topic of debate. When the X-ray spectrum can be described by a soft, thermal blackbody, the origin is usually attributed to the inner, hot regions of the accretion disc \citep[one exception is the neutron star surface;][]{gilf10}. In the case of BHXBs, this soft thermal component often dominates (the X-ray `soft state') and when it does not, the X-ray spectrum can usually be described by a hard power law \citep[the X-ray `hard state'; e.g.][]{bell10}. The inner, hot, possibly radiatively inefficient accretion  flow/`corona' is generally considered to produce this power law, due to Compton upscattering of soft photons on hot electrons \citep[see][ for a review]{gilf10}.

\cite*{market01} first proposed that optically thin synchrotron emission from the jet dominates the X-ray flux of the BHXB XTE J1118+480. This was based on broadband, radio to X-ray spectral modelling when the source was accreting at $\sim 10^{-3} L_{\rm Edd}$, where the hard power law at X-ray energies could instead be explained by optically thin jet emission extending from the optical regime. These and similar models taking into account the jet and Comptonization were developed to explain the broadband SEDs of BHXBs in the hard state, and it was shown that the synchrotron component probably produces some fraction of the X-ray flux in the hard state, but may not dominate \citep[e.g.][]{market03,market05,miglet07,maitet09,zdziet12,peerma12}.

More recently, empirical evidence for the jet producing the hard X-ray power law has been found in the BHXB XTE J1550--564. Here, the jet emission at optical/NIR frequencies was isolated from the emission from the accretion disc and companion star \citep{russet10}. This jet component was found to have a spectral index consistent with optically thin synchrotron emission during the fading hard state of its outburst in 2000. The spectral index between X-ray and NIR, the spectral index of the jet component measured at optical/NIR frequencies and the X-ray spectral index itself, were all consistent with the same value. The NIR emission from the jet was also found to be linearly proportional to the simultaneous X-ray flux, and the evolution of the broadband spectrum from optical to X-ray energies could be approximated by one single power law fading by one order of magnitude in flux. Since the optical/NIR optically thin emission originated in the jet \citep[see also][]{chatet11}, this implied that the NIR--X-ray power law originated in the optically thin emission from the jet synchrotron spectrum during the fading hard state, at a luminosity of $\sim (2 \times 10^{-4} $ -- $ 2 \times 10^{-3})$ $L_{\rm Edd}$. Initially in the hard state decay, when the jet was brightening and the X-ray luminosity was higher, the X-ray power law must have originated in a different component -- most likely Compton upscattering in the corona. Evidence to support this change in the source of the dominating X-ray emission comes from an excess in the X-ray light curve over an exponential decay at the epoch in which the jet would start to dominate, a slight change in the X-ray hardness, and an increase in the X-ray rms variability.

A similar, but subtly different result has come from multi-wavelength monitoring of the BHXB XTE J1752--223 \citep{russet12}. Here, the jet and accretion disc emission at optical/NIR frequencies were isolated during the decay of its 2009--2010 outburst using the same method as \cite{russet10}. The optical jet emission was found to rise and fade during the fading hard state, in a similar fashion to XTE J1550--564. In the case of XTE J1752--223, the jet was contemporaneous with a clear X-ray flare with the same morphology in the light curve, implying again a common emission mechanism. The NIR--X-ray spectral index was consistent with optically thin synchrotron emission, but the X-ray timing and spectral properties before and during the flare were the same within errors. This implies that either the jet and corona have very similar timing properties \citep[as appears to be the case;][]{caseet10}, or that the X-ray emission was not dominated by the jet. In both scenarios, the X-ray emitting component must have been well correlated with the jet emission seen at optical/NIR frequencies \citep{russet12}.

Some additional works have also shown evidence for two components producing the X-ray power law in the hard state, one of which could be the jet. This was demonstrated for the plateau state of GRS 1915+105 \citep[which is equivalent to the canonical hard state;][]{rodret08a,rodret08b} and for GX 339--4 and GRO J1655--40 at low luminosities in the hard state \citep{soboet11}. In H1743--322, a change in the X-ray emission mechanism was also implied by the emission becoming radiatively efficient above a critical X-ray luminosity \citep{coriet11}. Finally, two components were fitted to the X-ray to $\gamma$-ray spectrum of Cyg X--1 \citep{lauret11,zdziet12,zdziet13}, one which was approximately one order of magnitude (or more) fainter than the other at X-ray energies (see also Figs. 1 and 2).

The above evidence suggests that at some stages of a BHXB outburst, the majority of the X-ray flux originates in optically thin synchrotron emission from the compact jets in the system. Since it preferentially occurs at low luminosities ($\simlt 10^{-3}L_{\rm Edd}$), where count rates are usually too low for accurate spectral fitting, and since it seems to have a similar power law index and timing properties to the Comptonized corona \citep[see also][]{caseet10}, this synchrotron component has probably been largely overlooked so far. As has been demonstrated, the polarization properties of the synchrotron and Comptonization X-ray power laws are expected to differ, so differentiating between these two emission mechanisms will be possible with future X-ray polarimeters \citep[see also][]{peerma12}. For jet synchrotron emission, X-ray FLP of a few per cent could be detected, with variability in FLP on short timescales. If a BHXB were found to have a magnetic field as ordered as that of Cyg X--1, and if the synchrotron component dominated the X-ray emission in that source at a certain time, one could expected very high (up to $\sim 70$ per cent), transient X-ray FLP.

X-ray polarization capabilities on board new X-ray satellites have been proposed on numerous occasions, most recently on missions eventually dropped, such as the Gravity and Extreme Magnetism Small Explorer and the New Hard X-ray Mission \citep[GEMS and NHXM;][]{blacet10,taglet12}, but such facilities have yet to be approved and launched. However, promise of new X-ray polarimeters on stratospheric balloons in the near future \citep[e.g. PoGOLite and X-Calibur;][]{pearet13,guoet13} may provide the most sensitive polarimetric X-ray measurements of any astrophysical sources to date.

\begin{figure}
\centering
\includegraphics[width=8.3cm,angle=0]{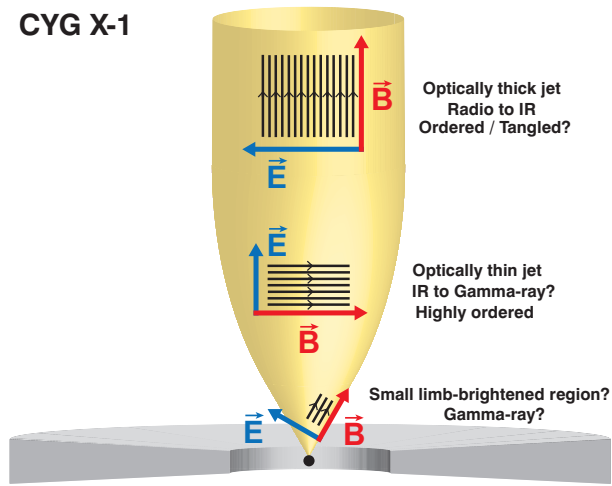}
\caption{A schematic diagram of the Cyg X--1 jet. This visual illustration assumes (i) a highly ordered magnetic field in the optically thin region near the jet base (as implied by the observations), (ii) an ordered or tangled field in the large scale jet (this can be tested using radio--mm polarimetry), and (iii) a jet opening angle of $60^{\circ}$, with the highly polarized MeV photons originating in a limb-brightened region with a magnetic field aligned with this angle (but PA rotation due to a steepening spectrum is more likely; see text).}
\end{figure}

\section{Conclusions}

We have presented new IR polarimetric data of Cyg X--1 using the 10.4-m Gran Telescopio Canarias and the 4.2-m William Herschel Telescope. These measurements have been combined with radio, optical, UV, X-ray and $\gamma$-ray flux and polarization data from the hard state. We use a simple phenomenological model to estimate the contribution of the jet, Comptonized corona and stellar companion at all frequencies. We find that our model is able to reproduce the observed flux spectrum, polarization spectrum and PA spectrum in the hard state self-consistently if the jet has a highly ordered magnetic field near its base, and dominates the MeV energies and the IR to radio regime. The model requires the magnetic field to be perpendicular to the axis of the resolved radio jet in order to reproduce the observed X-ray and IR PA. Our results imply that the historical X-ray polarization measured in Cyg X--1 \citep{longet80} was in fact due to synchrotron emission from the jet, and does not require the Comptonized corona to be polarized even though it dominates the X-ray flux. The magnetic field must be highly ordered and stable over several years in order to explain the detections of highly polarized $\gamma$-rays \citep{lauret11,jouret12}. Although interstellar dust is responsible for the majority of the optical and NIR $J$ and $H$-band polarization, the $K_{\rm S}$-band polarization is consistent with a contribution from the optically thin synchrotron power law from the inner regions of the jet, which extends to $\gamma$-ray energies. The observed position angle of the $\gamma$-ray polarization differs to the model by $\sim 60^{\circ}$. This is most likely due to the break or cut-off in the synchrotron spectrum residing at these energies, and this steepening of the spectrum can be associated with position angle shifts of $\sim 60^{\circ}$ according to numerical work.

Our mid-IR and NIR data from the soft state in 2013 do not provide evidence for synchrotron emission in the soft state. Although the $K_{\rm S}$-band polarization is slightly higher than expected from interstellar dust, the position angle is different in the soft state to the hard state (being consistent with the interstellar PA in the soft state), and the mid-IR FLP upper limit in the soft state is lower than the hard state model prediction.

The highly ordered magnetic field in the jet of Cyg X--1 is unprecedented, since other BHXBs and AGN tend to have predominantly tangled fields. However, Cyg X--1 is the first HMXB in which this has been well measured, and could imply a different field geometry in this type of accreting system. Our results could be confirmed and model parameters further constrained with more observations of Cyg X--1. The model predicts:
\begin{itemize}
\item A high level of polarization in the mid-IR in the hard state -- above $\sim 10$ per cent at 10--20 $\mu$m. We have demonstrated that this is detectable with the mid-IR CanariCam polarimetric instrument on the 10.4m Gran Telescopio Canarias, but the source was not in the hard state when our data were taken, so similar data acquired in the hard state would be very valuable. If the high, $\sim 10$ per cent FLP is detected at mid-IR frequencies, with the predicted PA of $\sim 160^{\circ}$, this would be a strong indication that the $\gamma$-ray polarization is robust and that the jet of Cyg X--1 indeed has a highly ordered magnetic field.
\item A shift in the PA by $90^{\circ}$ around the frequency of the jet break. The change in FLP is more smooth than this sharp change in PA, so this is a useful prediction that can be tested with polarimetry at frequencies just lower and just higher than the jet break. If the PA shift is this sharp it could even yield a more accurate jet break frequency than the flux spectrum and FLP can provide.
\item Radio FLP up to 10 per cent, if the magnetic field remains ordered down the length of the jet. More constraining radio polarimetric observations of Cyg X--1 can test whether the magnetic field structure changes as a function of distance along the jet.
\item A steeply increasing FLP at hard X-ray--$\gamma$-ray energies, accompanied by a shift in PA by $\sim 60^{\circ}$ to fit the $\gamma$-ray polarization measurements of INTEGRAL. Specifically, an increase from FLP $\sim 4$ per cent at 50 keV to $\sim 50$ per cent at 500 keV is inferred from the model. This could be tested with future X-ray polarimeters.
\end{itemize}

From observations of Cyg X--1 and other BHXBs, we predict that variable X-ray polarization from synchrotron emitting jets could be detected from accreting black holes by future X-ray polarimeters. Variable X-ray FLP of up to 10 per cent could be observed, opening up a new field of study that could be compared to models of jet production. Multi-wavelength polarization campaigns could greatly advance our understanding of how accretion in the strong gravitational fields close to black holes can result in the launching of relativistic jets.

\vspace{5mm}
\emph{Acknowledgements}.
We thank Jos\'e Acosta-Pulido and Antonio Pereyra for help with the LIRIS data reduction pipeline, Carlos \'Alvarez Iglesias for advice regarding the reduction of CanariCam polarimetric data and Gabriel P\'erez D\'iaz (Servicio MultiMedia, IAC) for the schematic illustration (Fig. 3). We would also like to thank Sera Markoff, Victoria Grinberg, Julien Malzac and Marion Cadolle Bel for conversations on the subject and in particular, Tom Maccarone for insightful discussions regarding the expected $\gamma$-ray emission from Comptonization of synchrotron photons. Based on observations made with the Gran Telescopio Canarias (GTC), installed
at the Spanish Observatorio del Roque de los Muchachos of the Instituto de Astrof\'isica de Canarias, on the island of La Palma, and scheduled observations made with the William Herschel Telescope (WHT) operated on the island of La Palma by the Isaac Newton Group in the Spanish Observatorio del Roque de Los Muchachos.  \textsc{pyraf} is a product of the Space Telescope Science Institute, which is operated by AURA for NASA. DMR acknowledges support from a Marie Curie Intra European Fellowship within the 7th European Community Framework Programme under contract no. IEF 274805. DMR and TS acknowledge support from the Spanish Ministry of  Science and Innovation (MICINN) under the grant AYA2010-18080.

\end{document}